\documentclass[12pt]{article}
\usepackage{amsbsy}

\newcommand{\sect}[1]{\setcounter{equation}{0}\section{#1}}
\newcommand{\eq}{\begin{equation}}
\newcommand{\eqa}{\begin{eqnarray}}
\newcommand{\en}{\end{equation}}
\newcommand{\ena}{\end{eqnarray}}
\newcommand{\enn}{\nonumber \end{equation}}

\def\sk{\vskip .4cm}
\def\noi{\noindent}
\def\om{\omega}
\def\al{\alpha}
\def\be{\beta}
\def\ga{\gamma}

\let \part\partial

\def\unquarto{{1 \over 4}}

\def\unmezzo{{1 \over 2}}
\def\epsi{\varepsilon}
\def\we{\wedge}

\def\de{\delta}

\def\part{\partial}

\def\sk{\vskip .4cm}

\def\noi{\noindent}

\def\X0{X^0}

\def\om{\omega}

\def\al{\alpha}
\def\ga{\gamma}

\def\unquarto{{1 \over 4}}
\def\unmezzo{{1 \over 2}}
\def\epsi{\varepsilon}
\def\epsibold{{\bf \epsilon}}

\def\we{\wedge}

\def\de{\delta}

\def\Rcal{{\cal R}}
\def\Ccal{{\cal C}}

\def\square{{\,\lower0.9pt\vbox{\hrule \hbox{\vrule height 0.2 cm
\hskip 0.2 cm \vrule height 0.2 cm}\hrule}\,}}

\def\epsilonbar{{\bar \epsilon}}

\def\westar{\we_\star}

\def\psibar{\bar \psi}
\def\chibar{\bar \chi}

\def\Om{\Omega}

\def\Sigmabar{\overline \Sigma}

\def\Rbold{{\bf R}}
\def\Ombold{{\bf \Om}}
\def\onebold{{\bf 1}}
\def\epsibold{\boldsymbol {\epsilon}}

\def\Gammabold{{\bf \Gamma}}

\begin{document}

\begin{titlepage}
%\rightline{DISTA-UPO/11}
%\rightline{hep-th/9509031}
%\rightline{November 2011} \vskip 2em
\begin{center}
{\Large \bf Chern-Simons supergravities, with a twist }
\\[3em]
{\large {\bf Leonardo Castellani} } \\ [2em] {\sl Dipartimento di Scienze e Innovazione Tecnologica
\\ INFN Gruppo collegato di Alessandria,\\Universit\`a del Piemonte Orientale,\\ Viale T. Michel 11,  15121 Alessandria, Italy
}\\ [4em]
\end{center}

\begin{abstract}

\vskip 0.2cm
We discuss noncommutative extensions of Chern-Simons (CS) supergravities in odd dimensions.
The example of $D=5$ CS supergravity, invariant under the gauge supergroup $SU(2,2|N)$,  is worked out  in detail. Its noncommutative version, with a $\star$-product associated to an abelian Drinfeld twist,  is found to exist only for $N=4$.

 \end{abstract}

\vskip 11cm \noi \hrule \vskip.2cm \noi {\small 
leonardo.castellani@mfn.unipmn.it}

\end{titlepage}

\newpage
\setcounter{page}{1}

\sect{Introduction}

Chern-Simons (CS) supergravities \cite{Troncoso1998,Zanelli2005,Zanelli2012} offer an interesting alternative to standard supergravities
for at least two reasons:

$\bullet$ ~~supersymmetry is realized as a {\sl gauge} symmetry, part of a gauge supergroup $G$ under which the CS  Lagrangian is invariant up to a total derivative. The superalgebra closes off-shell by construction.

$\bullet$ ~~the gauge supergroup contains the (anti)-De Sitter superalgebra, so that the theory
is translation-invariant and does not have dimensionful coupling constants. Group contraction
can be used to recover the Poincar\'e superalgebra, and the corresponding Poincar\'e supergravity.

Both these features can be relevant for a consistent  quantization of the theory \cite{Zanelli2005}. 
CS gravities and supergravities live only in odd dimensions $D=2n-1$, and contain,
besides the usual Einstein-Hilbert term and its supersymmetrization, also a cosmological term (in the uncontracted version)
and  higher powers of the curvature $2$-form $R$ up to order $n-1$. Note also that CS gravities are a particular example of Lovelock gravities \cite{Lovelock}, with at most second order field equations for the metric.

In this paper we present a noncommutative (NC) extension of five dimensional Chern-Simons supergravity. We find
that noncommutativity requires a particular value of $N$ in the gauge supergroup $SU(2,2|N)$, namely $N=4$,
for which the supergroup becomes nonsimple, with a central $U(1)$. 
The gauge potentials of $SU(2,2|4)$ are the fields appearing in the noncommutative action: the vielbein $V^a$, four gravitini $\psi_i$, the spin connection $\om^{ab}$, the $SU(4)$ gauge field $a^i_{~j}$, and an extra $U(1)$ gauge field $b$
(necessary for the NC extension). 

Physical motivations for studying (super)gravity on noncommutative spacetime, reformulated as a field theory on ordinary spacetime but with a deformed $\star$-product between fields, have been extensively discussed in the last two decades.
The possibility of encoding quantum properties in the texture of spacetime, and obtain deformations of gauge and gravity theories (invariant under deformed symmetries), is one among the interesting applications of noncommutative geometry to physics.
Comprehensive reviews can be found for example in ref.s 
\cite{Connes,Landi,madorebook,Castellani1,revDN,revSz1,book}. 

The $D=5$ NC Chern-Simons action studied here  is invariant under the full supergroup noncommutative transformations (or $\star$-transformations), in  particular under the four $\star$-supersymmetries. It provides an example of locally $\star$-supersymmetric noncommutative  theory in $D = 5$. (A $D=3$ locally supersymmetric theory was constructed in \cite{Cacciatori2002}, see also ref. \cite{AC2,Castellani2} for a $D=4$ noncommutative supergravity).

Four-dimensional noncommutative gravities
based on topological actions {\` a} la Mac-Dowell-Mansouri have been considered in \cite{Cham2002,Garcia2002,CZ,AC1,DRS}, and in \cite{Castellani2} for
$OSp(1|4)$ noncommutative supergravity.  One of the early works on higher dimensional Chern-Simons actions in noncommutative spaces is ref. \cite{CF1994}.

The paper is organized as follows. Section 2 briefly recalls the definition of Chern-Simons forms and some of their properties. In Section 3 
we discuss their noncommutative extensions. Section 4 deals with noncommutative $D=5$ CS supergravity,
and Section 5 contains some conclusions and outlook. In Appendix A 
we discuss in some detail the gauge variation of noncommutative CS forms, with the explicit example of $D=3$.
Finally in Appendix B we collect 
conventions and useful formulas for $D=5$ gamma matrix algebra.

\sect{Chern-Simons forms}

By definition a CS Lagrangian $L_{CS}^{(2n-1)}$ is a ($2n-1$)-form whose exterior derivative yields a gauge invariant $2n$-form.
In the present note we concentrate on the case
 \eq
 dL_{CS}^{(2n-1)} = STr(R^n)  \label{CSdef}
 \en
 where $R^n \equiv R \we R \we \cdots \we R $ ($n$ times),  the curvature $2$-form $R$ being defined 
as $R=d\Om - \Om \we \Om$, and $L_{CS}^{(2n-1)}$ contains (exterior
products of) the $G$ gauge potential one-form $\Om$ and its exterior derivative. The supertrace
$STr$ is taken on some representation of the supergroup $G$.

  Thus the CS action is related to a topological action in $2n$ dimensions  via Stokes theorem:
 \eq
  \int_{\part M} L_{CS} ^{(2n-1)}= \int_{M} Str(R^n)
 \en
 Gauge transformations are defined by
 \eq
 \de_\epsi \Om = d \epsi - \Om \epsi + \epsi \Om,~~~ \Rightarrow ~~~\de_\epsi R = - R \epsi + \epsi R \label{gaugetransf}
 \en
 so that $STr(R^n)$ is manifestly gauge invariant. Then (\ref{CSdef}) implies that the gauge variation of 
$L_{CS}^{(2n-1)}$  is closed, and hence locally exact:
 \eq
 \de_\epsi L_{CS} ^{(2n-1)}= d \alpha^{(2n-2)}(\Om, R, \epsi)
 \en
\noi We conclude that the Chern-Simons action is gauge invariant
 \eq
 \de_\epsi \int L_{CS}^{(2n-1)} = 0
  \en
  with suitable boundary conditions.
  
  The CS Lagrangian is given in terms of $\Om$ and  $d\Om$ (or $R$) by the following expressions \cite{Nakahara,EGH}:
   \eq
    L_{CS}^{(2n-1)}= n \int^1_0 STr[\Om(t d\Om-t^2 \Om^2)^{n-1} ] dt= n \int^1_0 t^{n-1} STr[\Om(R+(1-t)\Om^2)^{n-1} ] dt
     \en
     For example:
      \eqa
      & & L^{(3)}_{CS} = STr[R\Om + {1 \over 3} \Om^3] \\
      & & L^{(5)}_{CS} = STr[R^2\Om + \unmezzo R \Om^3 + {1 \over 10} \Om^5] \label{CS5}\\
        & & L^{(7)}_{CS} = STr[R^3\Om + {2 \over 5} R^2 \Om^3 + {1 \over 5} R \Om^2 R \Om +  {1 \over 5} R \Om^5
        +{1 \over 35} \Om^7]
         \ena
  
  \noi In the following  $L_{CS}^{(2n-1)}$ is always considered as a function of $\Om$ and $R$.
 Its gauge variation is easily computed using (\ref{gaugetransf}). For example:
   \eq
    \de_\epsi L^{(5)}_{CS} =  STr[R^2 d\epsi + \unmezzo R (d\epsi \Om^2 + \Om d\epsi \Om + \Om^2 d\epsi) +
       {1 \over 2} d\epsi \Om^4] 
 \en
  The general rule to obtain the variation is simple: just replace in $L_{CS}^{(2n-1)}$ each $\Om$ factor 
  in turn by $d \epsi$ (the terms with undifferentiated $\epsi$ cancel out because of the cyclicity of $STr$).
   Using now
   \eq
    d\Om = R + \Om \Om,    ~~~~~~ dR = \Om R - R \Om~~\rm{(Bianchi ~identity)}
     \en
  one recognizes that
 \eq
    \de_\epsi L^{(5)}_{CS} = d ~STr[R^2 \epsi + \unmezzo R (\epsi \Om^2 - \Om \epsi \Om + \Om^2 \epsi) +
       {1 \over 2} \epsi \Om^4] 
  \en
 This leads to the general recipe
  \eq
   \de_\epsi L_{CS}^{(2n-1)} = d ( j_\epsi L_{CS}^{(2n-1)}) \label{CSvariation}
   \en
 \noi where $j_\epsi$ is a contraction acting selectively on $\Om$, i.e.
  \eq
   j_\epsi \Om = \epsi,~~~   j_\epsi R = 0
    \en
    with the graded Leibniz rule $j_\epsi (\Om \Om) = j_\epsi (\Om) \Om - \Om  j_\epsi (\Om) = \epsi \Om - \Om \epsi$ etc. 
    
 \sect{Noncommutative CS actions}
 
The preceding discussion holds for a generic supergauge connection $\Om$, and relies only on the (graded) cyclicity of 
the supertrace. As such, it can be extended without effort to construct noncommutative Chern-Simons actions, where the noncommutativity is controlled by an abelian twist. This amounts to a deformation of the exterior product:
 \eqa
 & &  \tau \westar \tau' \equiv  \sum_{n=0}^\infty \left({i \over 2}\right)^n \theta^{A_1B_1} \cdots \theta^{A_nB_n}
   (\ell_{X_{A_1}} \cdots \ell_{{A_n}} \tau) \we  (\ell_{{B_1}} \cdots \ell_{{B_n}} \tau')  \nonumber \\
  & & ~~ = \tau \we \tau' + {i \over 2} \theta^{AB} (\ell_{A} \tau) \we (\ell_{B} \tau') + {1 \over 2!}  {\left( i \over 2 \right)^2} \theta^{A_1B_1} \theta^{A_2B_2}  (\ell_{{A_1}} \ell_{{A_2}} \tau) \we
 (\ell_{{B_1}} \ell_{{B_2}} \tau') + \cdots \nonumber \\
  \label{defwestar}
\ena
       \noi where $\theta^{AB}$ is a constant antisymmetric matrix, and $\ell_{A}$ are Lie derivatives along commuting
       vector fields $X_A$. This noncommutative product is associative due to $[X_A , X_B]=0$.  If  the vector fields $X_A$ are chosen to  coincide with the partial derivatives
$\partial_\mu$, and if $\tau$, $\tau'$ are $0$-forms, then $\tau\star\tau'$ reduces to the well-known Moyal-Groenewold product \cite{MoyalGroenewold}. A short review on twisted differential geometry can be found for example in \cite{AC1}.
\sk
However, the supertrace is not cyclic for twisted products. What is
(graded) cyclic is the {\it integrated} supertrace:
 \eq
 \int STr (\tau \westar \tau') = (-1)^{deg(\tau) deg(\tau')}
 \int STr (\tau' \westar \tau)  {\rm+~boundary~terms}
\en
This is sufficient to prove the $\star$-gauge invariance of the
noncommutative Chern-Simons action. Indeed if we denote
by $L^{(2n-1)}_{CS^*}$ the noncommutative Chern-Simons 
Lagrangian, obtained by substituting $\star$-exterior products
to ordinary exterior products, then
 \eq
 \de_\epsi^\star \int L^{(2n-1)}_{CS^*}=\int d ( j_\epsi L_{CS^*}^{(2n-1)}) =0
 \en
\noi for suitable boundary conditions. This is because the variation formula (\ref{CSvariation}) holds, but only under integration, also in the $\star$-deformed case, with $\star$-gauge transformations given by:
 \eq
 \de_\epsi^\star \Om = d \epsi - \Om \star \epsi + \epsi \star \Om,~~~ \Rightarrow ~~~\de_\epsi^\star R = - R \star \epsi + \epsi \star R \label{stargaugetransf}
 \en
For example the $D=5$ $\star$-Chern-Simons action reads
\eq
 \int L^{(5)}_{CS^*} = 
\int  STr[R \westar R \westar \Om + \unmezzo R \westar \Om \westar \Om \westar \Om + 
  {1 \over 10} \Om \westar \Om\westar \Om \westar \Om \westar \Om] \label{starCS5}
\en
\noi and  is invariant under the $\star$-gauge variations (\ref{stargaugetransf}).
  
 \sect{D=5 noncommutative CS supergravity}

The relevant supergroup for $D=5$ CS supergravity is $SU(2,2|N)$ (For a group-geometric construction of standard
$D=5$ supergravity see for ex. \cite{CDF}, p. 755). We begin by writing
 the noncommutative connection and curvature supermatrices.
The gauge connection $1$-form is given by:
 \eq
  \Ombold \equiv 
\left(
\begin{array}{cc}
  \Om^\al_{~\be} &  \psi_j^\al   \\
 -\psibar^i_\be &  A^i_{~j}   \\
\end{array}
\right), ~~~ \Om^\al_{~\be} \equiv (\unquarto \om^{ab} \ga_{ab} - {i \over 2} V^a \ga_a + {i \over 4} b I)^\al_{~\be}, ~~~ 
A^i_{~j} ={i \over N} b \de^i_j + a^i_{~j} 
    \label{Omdef}  
  \en
  where the bosonic $U(2,2)$ subgroup is gauged by the $1$-forms $\om^{ab}$ (spin connection), $V^a$ (vielbein) and $b$ ($U(1)$ gauge field); the antihermitian matrix-valued $1$-forms  $a^i_{~j}$ ($i,j=1...N$) gauge the $SU(N)$ bosonic subgroup; finally the $N$ gravitino $1$-form fields $ \psi_j$ gauge the $N$ supersymmetries. The Dirac conjugate is defined as $\psibar = \psi^\dagger \ga_0$. The NC connection coincides in fact with the commutative one: indeed for generic $N$ no extra fields are needed (contrary to the case of $D=4$ NC (super)gravity \cite{AC1,AC2}). This is due to the fact that the $D=5$ gamma matrices $\ga^a$, $\ga^{ab}$ and the identity matrix span a complete basis for $4 \times 4$ matrices.
  
The corresponding curvature supermatrix $2$-form is
 \eq
      \Rbold =  d \Ombold - \Ombold \westar \Ombold~
  \equiv  \left(
\begin{array}{cc}
   R + \psi_i \westar \psibar^i &  \Sigma_j    \\
 -\Sigmabar^i  &  F^i_{~j}+ \psibar^i \westar \psi_j   \\
\end{array}
\right) \label{Rdef}
       \en
\noi with
   \eqa
    & & R = d\Om - \Om \westar \Om \equiv \unquarto R^{ab} \ga_{ab} - {i \over 2} R^a \ga_a +{i\over 4} r I \label{defR}\\
    & & \Sigma_j = d \psi_j - \Om \westar  \psi_j - \psi_k  \westar A^k_{~j} \equiv D \psi_j \label{defSigma} \\
    & & \Sigmabar^i = d \psibar^i - \psibar^i \westar \Om  - A^i_{~k} \westar  \psibar^k \equiv D \psibar^i \\
    & & F^i_{~j}= dA^i_{~j} - A^i_{~k} \westar A^k_{~j}
    \ena
    \noi Immediate algebra yields the components of the $U(2,2)$ curvature $R$:
   \eqa
     & & R^{ab} = d \om^{ab} - \unmezzo \om_c^{~[a} \westar \om^{b]c}  + \unmezzo V^{[a} \westar V^{b]} \nonumber  \\
    & &~~~~~+ {i \over 4} \epsi_{cde}^{~~~ab} (\om^{cd} \westar V^e + V^e  \westar \om^{cd}) 
     - {i \over 4} ( \om^{ab} \westar b + b \westar \om^{ab})\\
     & & R^{a} = d V^{a} - \unmezzo (\om^a_{~b} \westar V^{b} - V^{b}  \westar \om^a_{~b})  \nonumber \\
     & &  ~~~~~~ + {i \over 8} \epsi^a_{~bcde}  \om^{bc} \westar \om^{de}   
    -{i \over 4} (V^a \westar b + b \westar V^a) \\
    & & r = db - {i \over 2} \om^{ab} \westar \om_{ab} - i V^a \westar V_a - {i \over 4} b \westar b
    \ena

 \noi A direct consequence of the curvature definition (\ref{Rdef}) is the Bianchi identity
  \eq
   d\Rbold = - \Rbold \westar \Ombold + \Ombold \westar  \Rbold
    \en
  \noi which becomes, on the supermatrix entries 
   \eqa
    & & dR = - R\westar \Om+\Om \westar R,~~dF=-F\westar A + A \westar F, \label{BianchiR}\\
    & & d\Sigma= -R\westar  \psi + \Om \westar \Sigma - \Sigma \westar  A + \psi \westar  F, \label{BianchiSigma}\\
    & & d\Sigmabar= - \Sigmabar \westar \Om + \psibar\westar  R - F \westar \psibar + A \westar  \Sigmabar
     \label{BianchiSigmabar}\ena
\subsection{ $SU(2,2|N)$ gauge transformations}

\noi The NC gauge transformations  (\ref{stargaugetransf})  close on the $\star$-Lie algebra:
  \eq
    [\de_{\epsibold_1},  \de_{\epsibold_2}] = \de_{\epsibold_1 \star \epsibold_2 - \epsibold_2 \star \epsibold_1}
     \en
 
\noi In the case at hand the $SU(2,2|N)$  gauge parameter is
given by the supermatrix
 \eq
  \epsibold \equiv 
\left(
\begin{array}{cc}
  \epsi^\al_{~\be} &  \epsilon_j^\al   \\
 -\epsilonbar^i_\be &  \eta^i_{~j}   \\
\end{array}
\right), ~~~ \epsi^\al_{~\be} \equiv (\unquarto \epsi^{ab} \ga_{ab} - {i \over 2} \epsi^a \ga_a + {i \over 4} \epsi I)^\al_{~\be}, ~~~ 
\eta^i_{~j} ={i \over N} \epsi \de^i_j + \epsi^i_{~j} 
\en
\sk
\noi and the NC gauge variations on the block entries of $\Om$ read
\eqa
& &   \de \Om = d \epsi - \Om \star \epsi + \epsi \star \Om + \psi_i \epsilonbar^i + \epsilon_i \psibar^i \\
& & \de \psi_i = d \epsilon_i - \Om \star \epsilon_i + \epsilon_j \star A^j_{~i} - \psi_j \star \eta^j_{~i} + \epsi \star \psi_i  \\
& & \de \psibar^i = d \epsilonbar^i +\epsilonbar^i \star \Om - A^i_{~j} \star \epsilonbar^j +  \eta^i_{~j} \star \psibar^j  - \psibar^i \star \epsi   \\
 & & \de A^i_{~j} = d  \eta^i_{~j} - A^i_{~k} \star  \eta^k_{~j} +  \eta^i_{~k} \star A^k_{~j} + \psibar^i \star \epsilonbar_j - \epsilonbar^i \star \psi_j
\ena
\noi On the $\Om$ component fields the gauge variations take the form
 \eqa
  & & \de \om^{ab} = d \epsi^{ab} -\om_c^{~[a} \star \epsi^{b]c} + \epsi_c^{~[a} \star \om^{b]c} + V^{[a} \star \epsi^{b]} + \epsi^{[b} \star V^{a]} + \unmezzo (\psibar \star \ga^{ab} \epsi - \epsilonbar \star \ga^{ab} \psi)   \nonumber \\
  & & ~~~~~~+ {i \over 4} \epsi^{ab}_{~~cde}(\om^{cd} \star \epsi^e - \epsi^e \star \om^{cd} + V^e \star \epsi^{cd} 
   - \epsi^{cd} \star V^e) -  {i \over 4} ( \om^{ab} \star \epsi - \epsi \star \om^{ab}) \nonumber \\
   & & ~~~~~~ - {i \over 4} (b \star \epsi^{ab} - \epsi^{ab} \star b)\\
   & & \de V^a = d \epsi^a - \unmezzo (\om^{ab} \star \epsi^b + \epsi^b \star \om^{ab}) +
    \unmezzo (V^b \star \epsi^{ab} + \epsi^{ab} \star V^b) - i(\psibar \star \ga^a \epsilon - \epsilonbar \star \ga^a \psi) \nonumber \\
    & & ~~~~~~+ {i\over 8} \epsi^a_{~bcde} ( \om^{bc} \star \epsi^{de} - \epsi^{de} \star \om^{bc})+
    {i \over 4} (\epsi^a \star b - b \star \epsi^a + \epsi \star V^a - V^a \star \epsi) \\
& & \de \psi_i = d \epsilon_i- \Om \star \epsilon_ i+ \epsilon_j \star A^j_{~i} + (\unquarto \epsi^{ab} \ga_{ab} - {i \over 2} \epsi^a \ga_a + {i \over 4} \epsi ) \star \psi_i - {i \over N} \psi_i\star \epsi - \psi_j \star  \epsi^j_{~i} \nonumber \\ \label{psivariation}
   \ena
   with $\Om$ and $A^j_{~i}$ given in (\ref{Omdef}). On the $b$ field we find, respectively from $\de \Om$ and 
$ \de A^i_{~j}$ 
 \eqa
  & & \de b = d \epsi -  i(\psibar \star \epsilon - \epsilonbar \star \psi) + {i \over 2} ( \epsi_{ab} \star \om^{ab} - \om^{ab} \star \epsi_{ab}) + i ( \epsi_a \star V^a - V^a \star \epsi_a) \nonumber \\
  & & ~~~~~~+ {i \over 4} ( \epsi \star b - b \star \epsi) \label{bvariation1} \\
 & & \de b = d \epsi - i(\psibar \star \epsilon - \epsilonbar \star \psi) + i  ( a^i_{~j} \star \epsi^j_{~i} - \epsi^j_{~i}
 \star  a^i_{~j})  \nonumber \\
  & & ~~~~~~+ {i \over N} ( \epsi \star b - b \star \epsi)  \label{bvariation2}
 \ena
 We see that the $b$ gauge variations are consistent only if $N=4$ (otherwise the variation of $b$ under $U(1)$ in  the second line of 
 (\ref{bvariation1}) would not agree with the second line of (\ref{bvariation2})). Thus a noncommutative extension 
of Chern-Simons $D=5$ supergravity exists only for $N=4$. In this case the supergroup $SU(2,2|4)$ is not simple
anymore and the $U(1)$ gauged by the $b$ field becomes a central extension. Note that 
in the commutative limit $\epsi \star b - b \star \epsi$ vanishes, and no condition on $N$ arises. 

Consider now the
$U(1)$ gauge variation of the gravitini, cf. (\ref{psivariation}):
 \eq
 \de \psi_i =  {i \over 4} \epsi \star \psi_i - {i \over N} \psi_i\star \epsi 
 \en
\noi For $N=4$ we see that in the commutative
limit the gravitini become uncharged with respect to this $U(1)$, but remain charged in the noncommutative setting.

 \subsection{The action}

Substituting $\Rbold$ and $\Ombold$ into the Chern-Simons action (\ref{starCS5}), we obtain
the noncommutative CS action invariant under the $SU(2,2|4)$ gauge variations of the preceding subsection. 
The result is
  \eq
  \int Str( L^{(5)}_{CS}) = \int L_{U(2,2)} + L_{A}+ L_{fermi} \label{starCSSG5}
    \en
 \noi with
 \eqa
 & &  L_{U(2,2)} =  Tr[ R \westar R \westar \Om + \unmezzo R \westar \Om^{\star 3} + {1 \over 10} \Om^{\star 5}] \\
 & & L_A ~~~~=  - Tr[F \westar F \westar A + \unmezzo F \westar A^{\star 3} + {1 \over 10} A^{\star 5} ]\\
& &  L_{fermi} = {3 \over 2}  \psibar \westar (R \westar \Sigma + \Sigma \westar F) + {3 \over 2} \Sigmabar \westar (R \westar \psi + 
\psi \westar F)  \nonumber \\
 & & ~~~~~~~~~~+  \psibar \westar \psi \westar (\psibar \westar \Sigma + \Sigmabar \westar \psi ) 
 \ena
 \noi where $\Om^{\star3} \equiv \Om \westar \Om \westar \Om$ etc.  In the commutative limit it reproduces 
the action discussed in refs. \cite{ChamD5,Troncoso1998,Zanelli2005}.The $b$ field kinetic term has two contributions, from the $R\westar R \westar \Om$ and the $F \westar F \westar A$ terms, and is proportional to:
 \eq
  ({1 \over 16} - {1 \over N^2}) (db \westar db \westar b)
 \en
\noi which vanishes for $N=4$, as in the commutative case. The essential difference is that  the gravitini retain here a nonvanishing  $U(1)$ charge.

Since the $\star$-product contains the imaginary unit in its definition, it is necessary to check the reality of the 
NC action. We first observe that the supermatrix connection $\Ombold$ satisfies a reality condition:
 \eq
   \Ombold^\dagger = - \Gammabold_0 \Ombold \Gammabold_0,~~ \Gammabold_0 \equiv 
   \left(
\begin{array}{cc}
  \ga_0 &  0   \\
 0  &  I   \\
\end{array}
\right)
    \en
    due to $\ga_{ab}$ being $\ga_0$ antihermitian (i.e. $\ga_{ab}^\dagger = - \ga_0 \ga_{ab} \ga_0$ etc), while $I$ and $\ga_a$ are $\ga_0$ -hermitian. Noting that $\Gammabold_0^2 = \onebold$, and that the $\Gammabold_0$-antihermiticity of  $\Ombold$  implies  $\Gammabold_0$-antihermiticity of $\Rbold$,  one easily proves that the NC action (\ref{starCS5}) or (\ref{starCSSG5}), multiplied by $i$, is real.
  
\sk
We can obtain a slightly more explicit form for
$\int  L_{U(2,2)} $ by splitting the $U(2,2)$ connection in its ``Lorentz + rest" parts as 
 \eq
 \Om =  \om + V, ~~~\om \equiv \unquarto \om^{ab} \ga_{ab},~~~V \equiv  - {i \over 2} V^a \ga_a + {i \over 4} b I
 \en
 and correspondingly the $U(2,2)$ curvature as
 \eq
 R = \Rcal + T - V \westar V, ~~~\Rcal \equiv d\om - \om \westar \om, ~~~T \equiv dV - \om \westar V - V \westar \om
 \en
 Then we find, after some integrations by parts and use of the Bianchi identities (\ref{BianchiR})-(\ref{BianchiSigmabar}):
 \eqa
 & &  \int L_{U(2,2)} = 3 \int Tr[  \Rcal \westar \Rcal \westar V - {2\over 3} \Rcal \westar V^{\star 3} + {1 \over 5} V^{\star 5} \nonumber \\
  & & ~~~~~~~~~~~+ \unmezzo (T \westar R + R \westar T) \westar V + {1 \over 3} T \westar T \westar V - \unmezzo T \westar V^{\star 3}] \nonumber \\
& & ~~~~~~~~~~~~ + \int Tr[\Rcal \westar \Rcal \westar \om + \unmezzo \Rcal \westar \om^{\star 3} + {1 \over 10} \om^{\star 5}] \label{LU22}
  \ena
\noi The last line is the integral of the noncommutative  Lorentz CS form $L_{Lorentz}$. Its integrated derivative gives the integrated  NC Pontryagin $6$-form:
  \eq
  \int d L_{Lorentz} = \int Tr[ \Rcal \westar \Rcal \westar \Rcal]
  \en
This $6$-form $Tr[ \Rcal \westar \Rcal \westar \Rcal]$ vanishes in the commutative limit, so that the ordinary $L_{Lorentz}$ is closed and its integral becomes a boundary term, vanishing for suitable boundary conditions. Thus the last line of (\ref{LU22}) is absent in the commutative limit, but gives a nonvanishing contribution in the NC action.

 \sect{Conclusions and outlook}

We have constructed a noncommutative version of Chern-Simons supergravity in five dimensions. The theory is invariant under the $\star$-gauge transformations of the  supergroup $SU(2,2|4)$. 
The geometric generalization of the Seiberg-Witten map \cite{SW} developed in refs. \cite{AC3,ACD2} can now be applied to
this NC action: the result is a classical higher-derivative deformation of  $SU(2,2|4)$ Chern-Simons  supergravity
in  $D=5$, where the zero order (in $\theta$) term coincides with the classical action of refs. \cite{ChamD5,Troncoso1998,Zanelli2005}. All higher-order (in $\theta$) corrections are separately gauge invariant under the ordinary $SU(2,2|4)$ gauge transformations. This work is in progress and will be reported in a separate paper.

\sk\sk
 \noi {\bf Acknowledgements}
 \sk
 \noi We thank Paolo Aschieri for useful comments on the manuscript.

\appendix

\sect{On the gauge variation of NC Chern-Simons forms}

We have argued in Section 3 that the integrated gauge variation of the NC Chern-Simons form vanishes
for suitable boundary conditions since it is equal to a surface term. 

This implies that the gauge variation is a total derivative $d \alpha (\Om, R, \epsi)$. 
That it must be so is also clear from the fact that all computations are similar to the commutative ones,
and these are based on the graded cyclicity of the (super)trace, which holds  in the noncommutative case
 up to total derivatives. 
 
 As an exercise, we determine here the explicit expression of $\al (\Om, R, \epsi)$
for the $D=3$ noncommutative Chern-Simons form. 

We note that the expression (\ref{defwestar}) for the $\star$-exterior product
can be rewritten in terms of a bidifferential operator $\Delta$:
\eqa & &
 \tau \westar \tau'  = \tau \we \tau' + {i\over 2} \theta^{AB} (l_A \tau) \we (l_B \tau') + {1 \over 2!}
 \left( {i \over 2} \right)^2
\theta^{A_1 B_1} \theta^{A_2 B_2} (l_{A_1} l_{A_2} \tau) \we
 (l_{B_1} l_{B_2} \tau') + \cdots \nonumber \\ & &
  ~~~~~~~ \equiv e^\Delta (\tau,\tau') \label{starproduct}
  \ena
  \noi where powers of  $\Delta$ are defined
  as
  \eqa
  & & \Delta^n (\tau,\tau') \equiv \left( {i \over 2} \right)^n
\theta^{A_1 B_1} \cdots \theta^{A_n B_n} (l_{A_1} \cdots l_{A_n}
  \tau) \we (l_{B_1} \cdots l_{B_n} \tau') \\
  & & \Delta^0 (\tau,\tau' ) \equiv \tau \we \tau'
  \ena
  \noi {} and $\ell_{A}$ are Lie derivatives along commuting
       vector fields $X_A$.   From the definition (\ref{starproduct}) one finds the following expression
for the $\star$-commutator
(when at least one of the forms is of even rank, so that $\tau \we \tau' = \tau' \we \tau$):

\eq
  \tau \westar \tau' - \tau' \westar \tau = 2 \theta^{AB} l_A \Big[ {\sinh \Delta \over
\Delta}(\tau, l_B \tau')\Big] , \label{comm1}
 \en
 \noi a generalization of the $\star$-commutator formulas of ref.s \cite{PX,ACD}. If $\tau \we \tau'$ is a form of maximal degree, the Lie derivative in (\ref{comm1}) can be
 replaced by $d i_A$,  where $i_A$ is the contraction along the vector field $X_A$. Moreover, under trace 
 the formula holds also for matrix-valued forms:
 
\eq
 Tr( \tau \westar \tau' - \tau' \westar \tau )= d ~Tr\Big(  2 \theta^{AB} i_A \Big[ {\sinh \Delta \over
\Delta}(\tau, l_B \tau')\Big]  \Big) \equiv d ~ Tr[ \Ccal(\tau,\tau')]  \label{comm2}
 \en
 \noi This is the relevant formula for cyclic reorderings in noncommutative Chern-Simons
 forms, which are traces of maximal (odd) degree forms: any splitting inside the trace involves always one even form.
  
 Consider now the $D=3$ noncommutative CS form
  \eq
    L^{(3)}_{CS^\star} = STr[R \westar \Om + {1 \over 3} \Om \westar \Om \westar \Om] 
     \en
  We can compute its variation under the gauge transformations (\ref{gaugetransf}) and find:
   \eqa
   & & \de  L^{(3)}_{CS^\star}= {1 \over 3} ~d  ~Tr [ 3~R \star \epsi + \epsi \star \Om \westar \Om - 
    \Om \star \epsi \westar \Om  +   \Om \westar \Om \star \epsi  \nonumber \\
    & & ~~~~~~~~~~~~~~~~~~~+ 2~ \Ccal (\epsi, R \westar \Om) + \Ccal(\epsi, \Om^{\star 3})
      - \Ccal (\Om \westar R, \epsi) \nonumber \\
          & & ~~~~~~~~~~~~~~~~~~~ - \Ccal(\Om, R \star \epsi) - \Ccal(\Om \star \epsi, R) - \Ccal(\Om \star \epsi \westar \Om, \Om) ]
   \ena
   In the commutative limit, the first line reduces to  $\de  L^{(3)}_{CS}$, while the second and third line vanish.

\sect{Gamma matrices in $D=5$}

We summarize in this Appendix our gamma matrix conventions in $D=5$.
\eqa
& & \eta_{ab} =(1,-1,-1,-1,-1),~~~\{\ga_a,\ga_b\}=2 \eta_{ab},~~~[\ga_a,\ga_b]=2 \ga_{ab}, \\
& & \ga_0\ga_1\ga_2\ga_3\ga_4=-1,~~~\epsi_{01234} =  \epsi^{01234}=1, \\
& & \ga_a^\dagger = \ga_0 \ga_a \ga_0,  \\
& & \ga_a^T =  C \ga_a C^{-1}, ~~~C^2 =-1,~~~C^\dagger=C^T =-C
\ena

\subsection{Useful identities}

\eqa
 & &\ga_a\ga_b= \ga_{ab}+\eta_{ab}\\
 & & \ga_{abc}  = {1 \over 2} \epsilon_{abcde} \ga^{de}\\
 & & \ga_{abcd}  = - \epsilon_{abcde} \ga^{e}\\
 & &\ga_{ab} \ga_c=\eta_{bc} \ga_a - \eta_{ac} \ga_b +{1 \over 2} \epsilon_{abcde} \ga^{de}\\
 & &\ga_c \ga_{ab} = \eta_{ac} \ga_b - \eta_{bc} \ga_a+{1 \over 2} \epsilon_{abcde} \ga^{de}\\
 & &\ga^{ab} \ga_{cd} = - \epsi^{ab}_{~~cde}\ga^e - 4 \de^{[a}_{[c} \ga^{b]}_{~~d]} - 2 \de^{ab}_{cd}
 \ena
\noi where
$\delta^{ab}_{cd} \equiv \frac{1}{2}(\delta^a_c\delta^b_d-\delta^b_c\delta^a_d)$, $\delta^{rse}_{abc} \equiv  {1 \over 3!} (\de^r_a \de^s_b \de^e_c$ + 5 terms), 
and indices antisymmetrization in square brackets has total weight $1$. 

 \subsection{Noncommutative $D=5$ Fierz identities}

\eq
 \psi \westar \chibar = - \unquarto (-1)^{pq} [(\chibar \westar \psi) I + (\chibar \westar  \ga^a \psi) \ga_a
 - \unmezzo (\chibar \westar  \ga^{ab} \psi) \ga_{ab} ] \label{Fierz}
 \en
 \noi where $\psi$ is a spinor $p$-form and $\chi$ is a spinor $q$-form.

\end{document}